\documentclass[aps,prl,preprint,showpacs]{revtex4}
\usepackage{amssymb}
\usepackage{amsmath}
\usepackage{epsfig}

\newcommand {\cn} {{\rm cn}}

\newcommand {\dn} {{\rm dn}}
\begin{document}

\title{Coexistence of Josephson oscillations and novel self-trapping regime 
in optical waveguide arrays}
\author {Ramaz Khomeriki${}^{1,3}$,  J\'er\^ome Leon${}^2$, Stefano
Ruffo${}^1$}
\affiliation {${\ }^{(1)}$Dipartimento di Energetica ``S. Stecco" and CSDC,
Universit\'a di Firenze and INFN, via s. Marta, 3, 50139 Firenze (Italy) \\ 
${\ }^{(2)}$ Laboratoire de Physique Th\'eorique et Astroparticules 
  CNRS-UMR5207, Universit\'e Montpellier 2, 34095 Montpellier (France)\\ 
${\ }^{(3)}$  Physics Department, Tbilisi State University, 0128
  Tbilisi (Georgia)}

\begin{abstract}   Considering the coherent nonlinear dynamics between two
weakly linked optical waveguide arrays, we find the first example of {\it
coexistence} of Josephson oscillations with a novel self-trapping regime.  This
macroscopic bistability is explained by proving analytically  the simultaneous
existence of symmetric, antisymmetric and asymmetric stationary  solutions of
the associated  Gross-Pitaevskii equation.  The effect is, moreover, 
illustrated and confirmed by numerical simulations. This property allows to
conceive an optical switch based on the variation of the refractive index of
the linking central waveguide.  

\end{abstract}

\maketitle

\paragraph{Introduction.}

Since its prediction \cite{joseph} in 1962 and the immediate experimental
verification \cite{anderson}, the Josephson effect has led to many
implementations in various branches of physics. This macroscopic quantum
tunneling effect, originally discovered in superconducting junctions, is caused
by the global phase coherence between electrons in the different layers.
Similar {\em Josephson oscillations}  have been discovered in liquid Helium
\cite{helium1,helium2} and in double layer quantum Hall systems
\cite{hall1,hall2}. 

The first realization of a bosonic Josephson junction has been recently
experimentally reported \cite{ober} for a Bose-Einstein condensate embedded in
a macroscopic double well potential. The difference of the latter from the
ordinary Josephson junction behavior is that the oscillations of population
imbalance are suppressed for high imbalance values and a self-trapping regime
emerges \cite{smerzi1,smerzi2}. The optical realization of a bosonic junction 
had been theoretically proposed much earlier by Jensen \cite{jensen}  who
considered  light power oscillations in two coupled nonlinear waveguides. 

In order to describe the macroscopic tunneling effect in bosonic junctions, the
nonlinear dynamics is usually mapped to a simpler system characterized by two
degrees of freedom (population imbalance and phase difference) while the
nonlinear properties of  the wave function within the single well are
neglected. In this approach the symmetric and antisymmetric stationary
solutions of the Gross-Pitaevskii equation \cite{gros} are used as a basis to
build a global wave function \cite{kivshar1,ananikian}. In the weakly nonlinear
limit this description is fully valid because those symmetric and antisymmetric
functions are the only solutions of the equation. For higher nonlinearities an
asymmetric stationary solution appears, which represents a novel self-trapping
state \cite{alberto}.

\begin{figure}[ht]
\epsfig{file=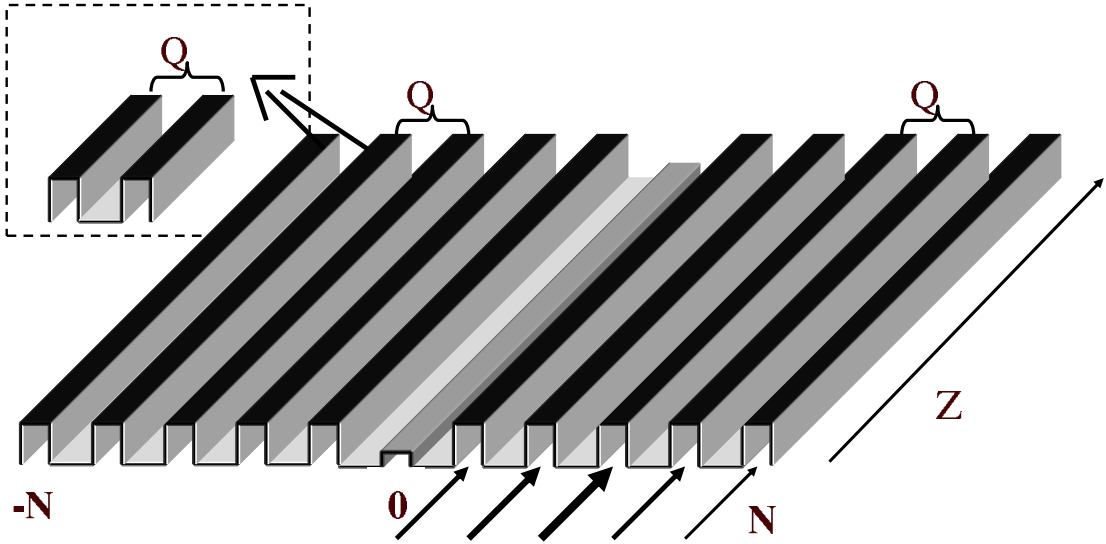,width=0.9\linewidth}
\caption{Scheme of the proposed setup of two weakly linked waveguide arrays.
The refractive index of the central waveguide is smaller than the indices of
the waveguides in the two arrays. The inset displays the elementary cell with
the coupling constant $Q$.}\label{fig:array} \end{figure}

In this Letter, we consider the two weakly linked optical waveguide arrays
represented in Fig.~\ref{fig:array}. When light is mainly injected in one array
(e.g. the right one, as shown in Fig.\ref{fig:array}), we find a wide light
intensity range where  it can either remain trapped in this array, or swing
periodically from right to left and back. The switching from one state to the
other is determined by a slight variation of the refractive index of the
central linking waveguide denoted with the index 0. The {\it coexistence} of
the oscillatory and self-trapping regimes corresponds to the simultaneous
presence of Josephson oscillation states and asymmetric solutions of the
Gross-Pitaevskii equation.   This is in contrast with all known behaviors of
bosonic Josephson junctions, where the existence of oscillatory and
self-trapping regimes is uniquely determined by the parameters of the system.
Our theoretical result is expected to have a straightforward experimental
realization in waveguide arrays and  in Bose-Einstein condensates with double
well potential.

\paragraph{The model.} 

Waveguide arrays are particularly convenient for a direct observation of 
nonlinear effects, because the longitudinal dimension $z$ plays the role of
time. With an intensity-dependent refractive index (optical Kerr effect),
waveguide arrays become soliton generators, as experimentally demonstrated in
\cite{eisenberg,morandotti,mandelik,yuri,fleisher1,fleisher2,assanto1}. 

The model of an array of adjacent waveguides coupled by radiation power
exchange is the Discrete NonLinear Schr\"odinger equation
(DNLS)~\cite{christ-joseph,mark} and  nonlinearity then manifests itself by a
self-modulation of the input signal (injected radiation). It reads
\begin{equation}
i\frac{\partial \psi_j}{\partial z}+\frac{\omega}
{c}n_j\psi_j+Q\bigl(\psi_{j+1}+\psi_{j-1}-2\psi_{j}\bigr)+ |\psi_j|^2\psi_j=0,
 \label{1} \end{equation}
where waveguides discrete positions are labelled by the index $j$  ($-N\leq j
\leq N$), and complex light amplitudes $\psi_j$ are normalized such as to fix
the onsite nonlinearity  to the unitary value. The linear refractive index
$n_j$ is set to $n$ for all  $j\ne 0$, and to $n_0<n$ for $j=0$. The coupling
constant between two adjacent waveguides is $Q$ and $\omega$ and $c$ are the
frequency and the velocity of the injected light, respectively. Last, we assume
vanishing boundary conditions $\psi_{N+1}=\psi_{-N-1}=0$ in order to mimick a
strongly evanescent field outside the waveguides.

The above equation, written for two waveguides (elementary cell in the inset of
Fig.\ref{fig:array}), straightforwardly reduces to the one considered by
Jensen~\cite{jensen}. In that case the resulting dynamics is given in terms of
Jacobi elliptic functions ~\cite{jensen,smerzi2} and can be described as
follows. For a beam of small intensity, light tunnels from one waveguide to the
other and then back, inducing Josephson oscillations. Indeed, the elementary
cell is similar to a single bosonic Josephson junction as demonstrated
in~\cite{smerzi1,smerzi2} and experimentally observed in Bose-Einstein
condensates~\cite{ober}. Increasing the injected beam intensity above some
critical value, light becomes self-trapped and does not tunnel to the other
waveguide, which constitutes the difference between a bosonic Josephson
junction and its superconducting analogue.

\paragraph{Numerical simulations.} 

We demonstrate now by numerical simulations of model~\eqref{1} that  for the
device of Fig.\ref{fig:array}, the two regimes, namely Josephson oscillations
and self-trapping, coexist for a given injected beam intensity and given
parameter values. The switch from one state to the other  is achieved, for
instance, by a tiny local variation of the refractive index of  the central
waveguide.

\begin{figure}[ht]
\epsfig{file=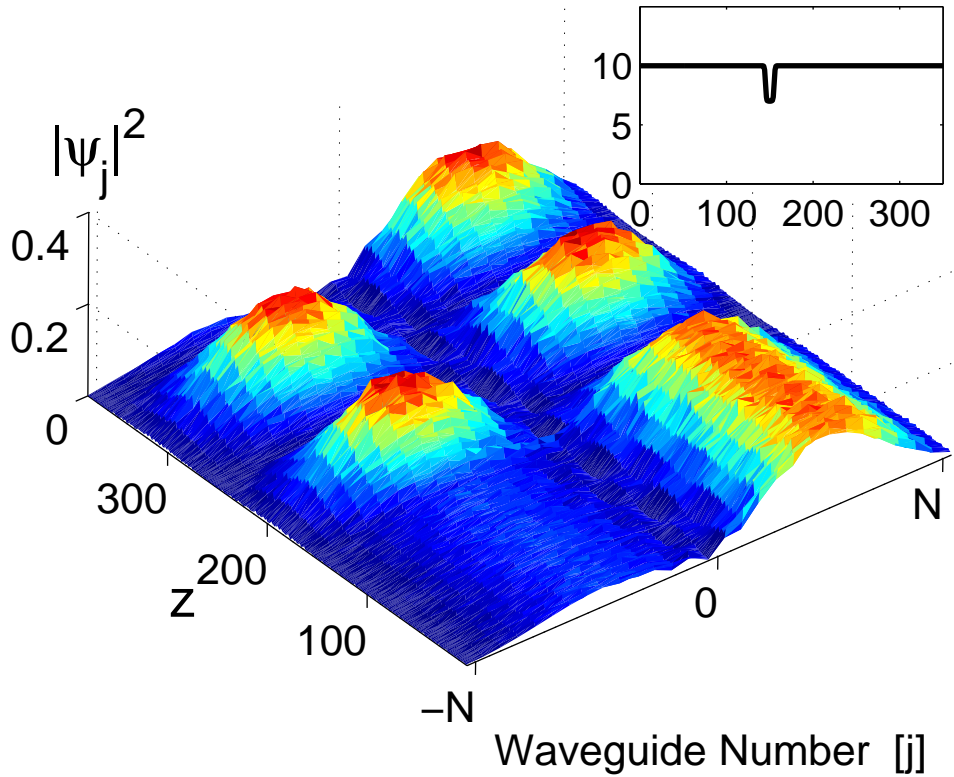,width=0.9\linewidth}
\caption{Numerical simulation of the DNLS equation~\eqref{1} with $z=0$ initial
conditions~\eqref{input} and parameters~\eqref{param}.  By a slight local
variation at $z=150$ of the refractive index of the central waveguide
(represented in the inset in terms of the relative barrier height $V_0$), the
regime switches from self-trapping to Josephson oscillations. The injected
total power is $P_t=\sum|\psi_j|^2=1.7$.} 
\label{fig:oscpin}\end{figure}

Let us choose the following values for the parameters in equation~\eqref{1}: 
\begin{equation}\label{param}
N=14,\quad Q=16,\quad V_0\equiv\omega(n-n_0)/2c=10, 
\end{equation}
together with the following input light envelope
\begin{eqnarray}\label{input}
\psi_j(0)=0.4\,\sin\bigl[(j-N-1)/5.5\bigr],\quad j=1,\cdots N, \\
\psi_j(0)=0.2\,\sin\bigl[(j+N+1)/5.5\bigr],\quad j=-N,\cdots 0, \nonumber
\end{eqnarray}
which represents a beam mostly sent into the right waveguide array.
Figure~\ref{fig:oscpin} displays the result of our numerical simulations. While
the relative refractive index $V_0$ of the central waveguide is kept constant,
the power injected initially into the right part of the array remains
self-trapped, as shown in Fig.\ref{fig:oscpin} up to $z=150$. A local variation
of $V_0$ at $z=150$, as drawn in the inset of Fig.\ref{fig:oscpin}, makes the
self-trapping state bifurcate to a regime of Josephson oscillations, which then
remains stable. 

\begin{figure}[t]
\epsfig{file=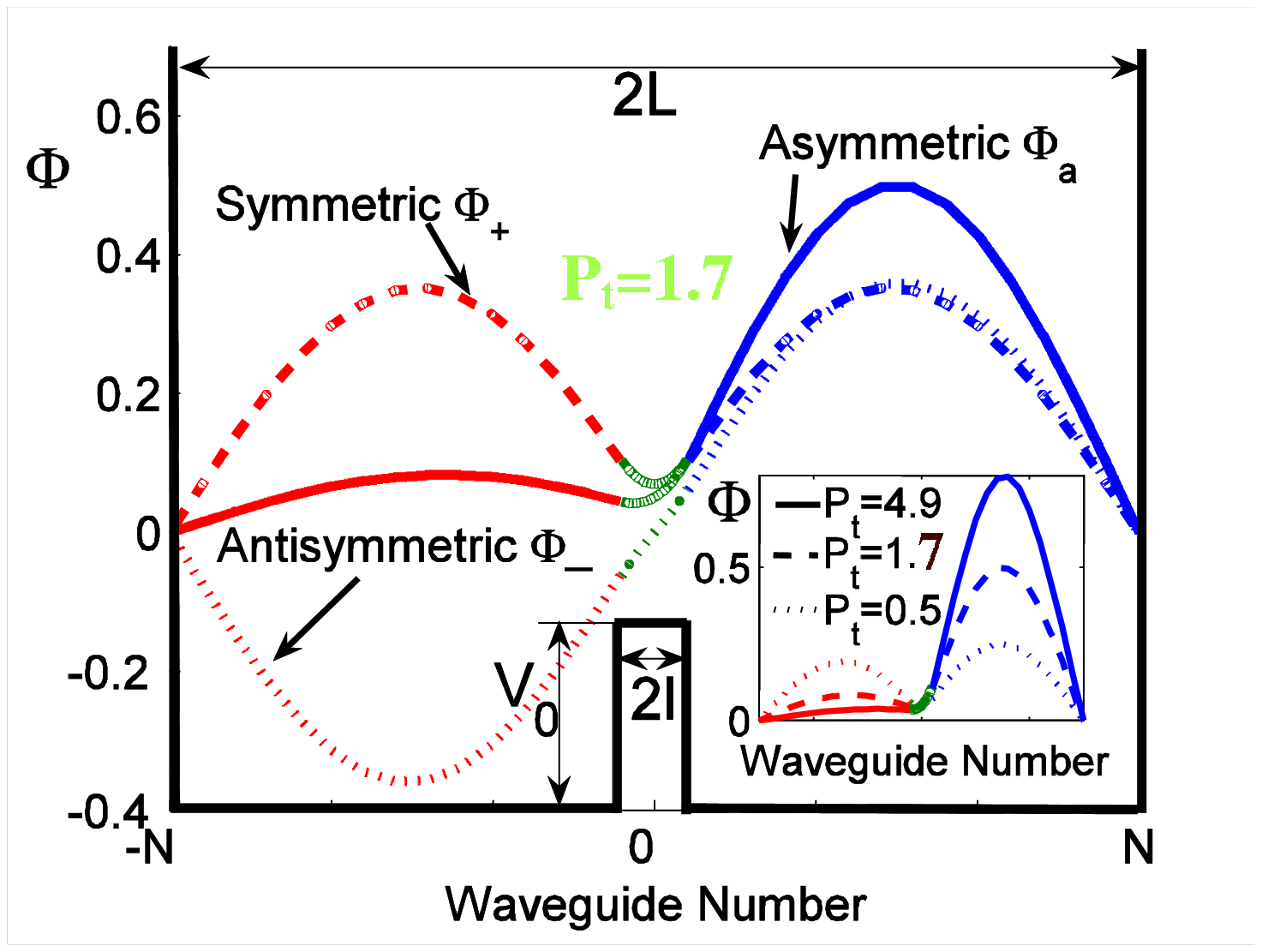,width=0.9\linewidth}
\caption{Plot of the double well potential for the continuous model \eqref{3}:
$2L$ is the well width, $V_0$ and $2l$ are barrier height and width. The curves
are the plots of different types of solutions for the total power $P_t=1.7$. 
The inset shows the form of the asymmetric solution for different values of the
total power.} \label{fig:0.25} \end{figure}

\paragraph{Theory.}
We shall now interpret these results in terms of the continuum limit of
model~\eqref{1}, which, after an  appropriate phase shift, reads as the
Gross-Pitaevskii equation 
\begin{equation}
i\frac{\partial \psi}{\partial z}+\frac{\partial^2 \psi}{\partial x^2}-V(x)
\psi+|\psi|^2\psi=0. \label{3}
\end{equation}   
The wavefunction $\psi(z,x)$ depends on the spatial continuous variables $z$
and  $x=j/\sqrt{Q}$.  $V(x)$ is a double well potential with a width
$2L=(2N+2)/\sqrt{Q}$ and is represented in Fig.~\ref{fig:0.25}. The potential
barrier of the double well potential has height $V_0$ and width
$2l=1/\sqrt{Q}$. In the region of the barrier we assume, for technical
simplification, that the Schr\"odinger equation can be treated in the linear
approximation, while numerical simulations are performed with a fully nonlinear
array. The stationary solution of~\eqref{3} are sought as $\psi(z,x)=\Phi(x)
\exp(-i\beta z)$, where the real-valued function $\Phi(x)$ is found in terms of
Jacobi elliptic functions as follows~\cite{book}:
\begin{align}
-l<x<-L\ :\ &\Phi=B\ \cn [\gamma_B(x+L)-{\mathbb K}(k_B),k_B],  \nonumber\\
-l<x<l\ :\ &\Phi=a e^{\lambda x}+ b e^{-\lambda x},
\label{4}\\
l<x<L\ :\ &\Phi=A\ \cn [\gamma_A(x-L)+{\mathbb K}(k_A),k_A] ,\nonumber
\end{align}
with the constants
\begin{equation}
\gamma_A=\sqrt{A^2+\beta}, \quad \gamma_B=\sqrt{B^2+\beta}, \quad
\lambda=\sqrt{V-\beta},
\end{equation}
${\mathbb K}$ denotes the complete elliptic integral of the first kind and the
moduli are
\begin{equation}
k_A^2=\frac{A^2}{2(A^2+\beta)}, \quad k_B^2=\frac{B^2}{2(B^2+\beta)}.
\end{equation}
The above solutions are given in terms of five parameters ($A$, $B$, $a$, $b$,
$\beta$). As the continuity conditions at points $x=\pm l$ provide four
relations, the total injected power $P_t=\int|\psi|^2dx$ completely determines
the solutions. 

In the weakly nonlinear limit $P_t\ll 1$, the solutions are either symmetric or
antisymmetric. The even solution $\Phi_+(x)$ corresponds to $A=B$ in~\eqref{4} 
when the solution in the barrier region is $2a\cosh(\lambda x)$. The odd
solution $\Phi_-(x)$ corresponds to $A=-B$ with central solution
$2a\sinh(\lambda x)$. For higher powers, namely above a threshold value, an
asymmetric solution $\Phi_a(x)$ also exists for which $A\neq \pm B$. These
analytical solutions are represented in Fig.~\ref{fig:0.25}.

Using the symmetric and antisymmetric basic solutions, one can build a
variational anzatz following~\cite{kivshar1,ananikian} by seeking the solution
$\psi(z,x)$ under the form
\begin{eqnarray}
\psi(z,x)=\psi_1(z)\Phi_1(x)+\psi_2(z)\Phi_2(x), \label{5} \\
\Phi_1=\frac1{\sqrt{2}}(\Phi_+ +\Phi_-), \quad \Phi_2=
\frac1{\sqrt{2}}(\Phi_+ -\Phi_-). \nonumber
\end{eqnarray}
The functions $|\psi_1|^2$ and $|\psi_2|^2$ are interpreted as the  {\em
probabilities} to find the light localized on the left or right array.  By
construction, the overlap of $\Phi_1$ with $\Phi_2$ is negligible, namely
\begin{equation}\label{overlap}
\int \Phi_1\Phi_2dx\ll
\int dx\ \Phi_1^2=\int dx\ \Phi_2^2,
\end{equation}
where the integrals run on $x\in[-L,L]$.
Consequently, the projection of  the Gross-Pitaevskii equation  \eqref{3} on
$\Phi_1$ and $\Phi_2$ provides the coupled mode equations \cite{jensen,smerzi1}
\begin{eqnarray}
i\frac {\partial \psi_1}{\partial z}-E_1\psi_1+ D|\psi_1|^2\psi_1=r\psi_{2},
\nonumber \\
i\frac {\partial \psi_2}{\partial z}-E_2\psi_2+ D|\psi_2|^2\psi_2=r\psi_{1},
\label{2} 
\end{eqnarray}
with coupling constant $r$ and nonlinearity parameter $D$ defined by
\begin{equation*}
r=
\frac {\int\left[(\partial_x\Phi_1)(\partial_x\Phi_2)+V\Phi_1\Phi_2\right]dx}
{\int\Phi_1^2\,dx}, \quad
D=\frac {\int\Phi_1^4\,dx}{\int\Phi_1^2\,dx}. \label{6}
\end{equation*}
The linear levels $E_1$ and $E_2$ are given by
\begin{equation}
E_n=\frac {\int\left[(\partial_x\Phi_n)^2+V\Phi_n^2\right]dx}
{\int\Phi_n^2\,dx},\quad n=1,2,
\end{equation}
and turn out to be equal thanks to \eqref{overlap}. As a consequence they can
be absorbed in a common phase in \eqref{2}.

An explicit solution of \eqref{2}  in terms of Jacobi elliptic functions  has
been found in \cite{jensen} and used in Bose-Einstein condensates in
\cite{smerzi2}. It has a simple form when all the power is initially injected
into one array, say $|\psi_1(0)|=1$, $|\psi_2(0)|=0$. Solutions have different
behavior depending on the value of $D$ being below or above the threshold value
$4r$
\begin{align}
D<4r\ :&\ |\psi_1(z)|^2=\frac12\left[1+\cn (2rz, \frac D{4r})\right],
\label{7a}\\
D>4r\ :&\ |\psi_1(z)|^2=\frac12\left[1+\dn (\frac D2z, \frac{4r}D)\right].
\label{7-bis}\end{align}
with $|\psi_2|^2=1-|\psi_1|^2$.
Solution \eqref{7a} describes an oscillation of light intensity between
the left and the right array (Josephson regime), since $|\psi_1|$ oscillates 
around the value $0$. The period of this oscillation is
\begin{equation}
T=2{\mathbb K}(D/4r)/r. \label{8}
\end{equation}
For $D>4r$ solution \eqref{7-bis} oscillates around the value 
$1/2+1/4(1+\sqrt{1-(4r/D)^2})$ with a period
\begin{equation}
T'=8{\mathbb K}(4r/D)/D, \label{8a}
\end{equation}
which shows that the system is in the self-trapping regime. Such simple
explicit formulas provide two important informations on the response of the
system to irradiation of one array, namely the period of the Josephson
oscillations and the threshold (in terms of parameter $D$) above which we
expect a bifurcation to a self-trapping regime.

\begin{figure}[t]
\epsfig{file=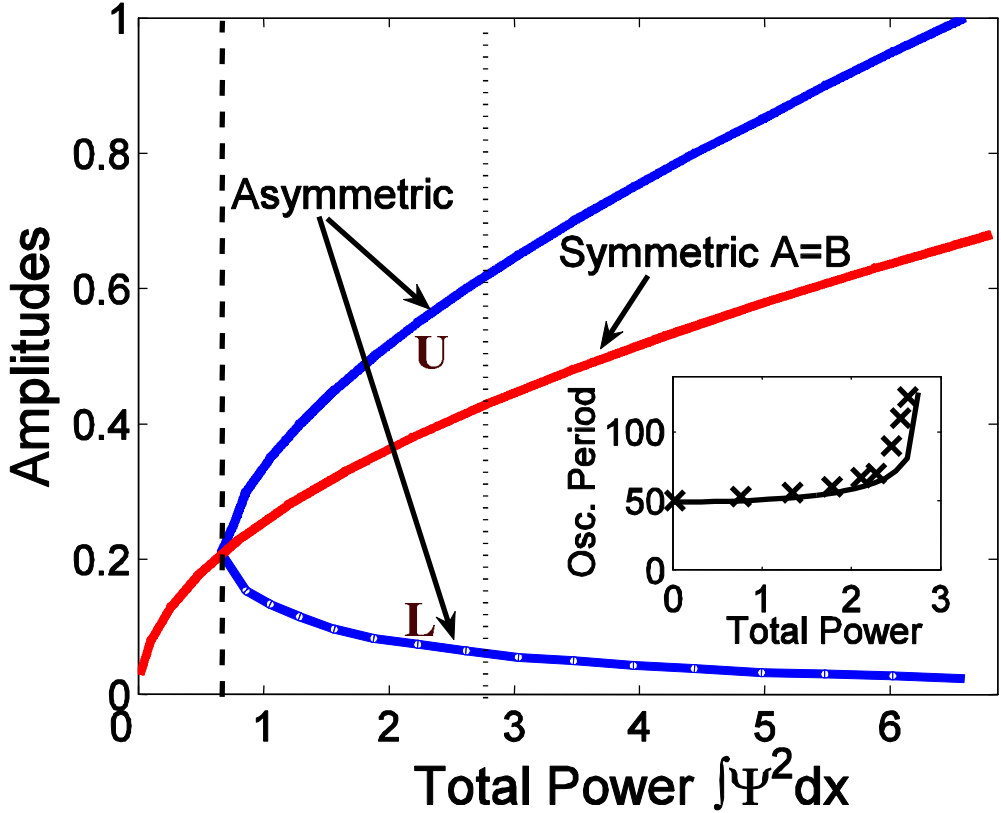,width=0.9\linewidth}
\caption{Main plot: dependence of the amplitudes (maximum values on
$x\in[-L,L]$) of the symmetric and asymmetric solutions \eqref{4} on the total
power (we have not reported the amplitude of the antisymmetric solution which
almost superposes to the one of the symmetric). The inset displays the period
of the Josephson oscillations for various values of the injected power: the
full line corresponds to the analytical expression \eqref{8}, the crosses are
results of numerical simulations.} \label{fig:bist} \end{figure}

However, the picture of possible solutions is not yet complete. We discuss here
the appearance of a further asymmetric solution when the injected power exceeds
a smaller threshold value. To be definite, we restrict to the parameter values
which follow from \eqref{param}: the width of the double well potential is
$2L=7.5$, the barrier width is $2l=0.25$  and its height is $V_0=10$. We derive
the complete set of solutions \eqref{4} and display the dependence of their
amplitudes on the total power $P_t=\int|\psi|^2dx$ in the main plot of
Fig.\ref{fig:bist}. Below the smaller threshold value $P_t=0.76$ (showed by the
vertical bold dashed line) only the symmetric and antisymmetric solutions exist
and their amplitudes almost superpose. At this threshold value a new solution
appears which is asymmetric and the amplitudes in the two arrays are
represented by the upper ($U$) and lower ($L$) branch in  Fig.~\ref{fig:bist}.
The existence of this solution is at the basis of a  novel self-trapping regime
as we explain below.  At a bigger threshold value (indicated by the vertical
dotted line)  solution \eqref{5} changes from the Josephson regime to the usual
self-trapping  one described above. 

In the region of injected power where the asymmetric solution coexists with 
the symmetric and asymmetric stationary solutions one can induce flipping  of
one to the other by varying the height of the barrier as shown in  Fig.
\ref{fig:oscpin}. It is likely that one can induce such flips by  other methods
(e.g. by the variation of the profile of the injected power).

Going back to Josephson oscillations related to the symmetric and
antisymmetric  solutions, a simple striking numerical check is provided by the
inset of  Fig.~\ref{fig:bist}, where we plot the period of the oscillation
against  the injected power. Numerical data, which are obtained with the fully 
discrete (and fully nonlinear) model \eqref{1} for parameters  in
\eqref{param}, compare extremely well with \eqref{8}. 

In conclusion a new coherent state in weakly linked waveguide arrays has been
discovered. This coherent state has the property of being bistable and one can
easily switch from oscillatory to self-trapping regimes and back. This
nontrivial behavior may have interesting applications in various weakly linked
extended systems, such as Bose-Einstein condensates or Josephson junctions
arrays, which deserve further studies.

\paragraph{Acknowledgements.} We would like to thank F.T. Arecchi, 
A. Montina and S. Wimberger for useful discussions. R. Kh. acknowledges support by Marie-Curie international incoming fellowship award (contract  No MIF1-CT-2005-021328) and NATO grant No FEL.RIG.980767. We also acknowledge financial support under the PRIN05 grant on {\it Dynamics and thermodynamics of systems with long-range interactions}.


\begin{thebibliography}{99}

\bibitem{joseph} B. D. Josephson, Phys. Lett. {\bf 1}, 251 (1962).
\bibitem{anderson}P.L. Anderson, J.W. Rowell, Phys. Rev. Lett. {\bf 10}, 230
(1963).
\bibitem{helium1} S.V. Pereverzev, A. Loshak, S. Backhaus, J. C. Davis, R. E.
Packard, Nature, {\bf 388}, 449 (1997).
\bibitem{helium2} K. Sukhatme, Y. Mukharsky, T. Chui, D. Pearson, Nature, {\bf
411}, 280 (2001).
\bibitem{hall1} I.B. Spielman et. al., 
Phys. Rev. Lett., {\bf 84}, 5808, (2000), {\em ibid}.
{\bf 87}, 036803 (2001).
\bibitem{hall2} M.M. Fogler, F. Wilczek, Phys. Rev. Lett., {\bf 86}, 1833,
(2001).
\bibitem{ober} M. Albiez, R. Gati, J. Folling, S. Hunsmann, M. Cristiani, M. K.
Oberthaler, Phys. Rev. Lett., {\bf 95}, 010402 (2005).
\bibitem{smerzi1} A. Smerzi, S. Fantoni, S. Giovanazzi, S. R. Shenoy, Phys.
Rev. Lett., {\bf 79}, 4950 (1997).
\bibitem{smerzi2} S. Raghavan, A. Smerzi, S. Fantoni, S. R. Shenoy, Phys. Rev.
A, {\bf 59}, 620 (1999).
\bibitem{jensen} S.M. Jensen, IEEE J. Quantum Electron. {\bf 18}, 1580, (1982). 
\bibitem{gros} L. P. Pitaevskii, Sov. Phys. JETP, {\bf 13}, 451, (1961); E. P.
Gross, Nuovo Cimento, {\bf 20}, 454, (1961); J. Math. Phys., {\bf 4}, 195,
(1963).
\bibitem{kivshar1}E. A. Ostrovskaya et. al., Phys. Rev. A, {\bf 61}, 031601(R), 
(2000).
\bibitem{ananikian} D. Ananikian, T. Bergeman, Phys. Rev. A, {\bf 73}, 013604,
(2006).
\bibitem{alberto} A. Montina, F.T. Arecchi, Phys. Rev. A, {\bf 66}, 013605,
(2002).
\bibitem{eisenberg}  H.S. Eisenberg et. al., Phys. Rev. Lett., {\bf 81}, 3383, 
(1998). 
\bibitem{morandotti}  R. Morandotti, H.S. Eisenberg, Y. Silberberg, M. Sorel,
J.S. Aitchison, Phys. Rev. Lett., {\bf 86}, 3296, (2001).
\bibitem{mandelik} D. Mandelik et. al., Phys. Rev. Lett., {\bf 90}, 053902, 
(2003); Phys. Rev. Lett., {\bf 92}, 093904, (2004).
\bibitem{yuri}A.A. Sukhorukov, D. Neshev, W. Krolikowski, Y.S. Kivshar,
Phys. Rev. Lett., {\bf 92}, 093901, (2004).
\bibitem{fleisher1} J.W. Fleischer et al., Phys. Rev. Lett., {\bf 90}, 023902,
(2003).
\bibitem{fleisher2} J.W. Fleischer et al., Nature, {\bf 422}, 147, (2003).
\bibitem{assanto1} A. Fratalocchi, G. Assanto, Phys. Rev. E {\bf 73}, 046603 
(2006); A. Fratalocchi et. al.,  Opt. Express, {\bf 13}, 1808, (2005). 
\bibitem{christ-joseph} D.N. Christodoulides, R.I. Joseph, Optics Lett., {\bf
13}, 794, (1988).
\bibitem{mark} M.J. Ablowitz, Z.H. Musslimani, Phys. Rev. Lett. {\bf 87},
254102, (2001); Phys. Rev. E, {\bf 65}, 056618, (2002).
\bibitem{book} P.F. Byrd, M.D. Friedman, {\it Handbook of elliptic integrals
for engineers and physicists}, Springer (Berlin 1954).

 



\end{thebibliography}
\end{document}